\documentclass{appolb}
\usepackage{graphicx}

\begin{document}
\title{Transition from ideal to viscous Mach cones in BAMPS%
\thanks{Presented at excited QCD 2012 in Peniche, Portugal}%
}
\author{Ioannis Bouras, Andrej El, Oliver Fochler, Carsten Greiner
\address{Institut f\"ur Theoretische Physik, Johann Wolfgang Goethe-Universit\"at\\
Max-von-Laue-Str.\ 1, D-60438 Frankfurt am Main, Germany}\\
Harri Niemi
\address{Department of Physics, P.O. Box 35 (YFL), FI-40014 University of
Jyv\"askyl\"a, Finland}
\\
Zhe Xu
\address{Department of Physics, Tsinghua University, Beijing 100084, China}
}
\maketitle
\begin{abstract}
We investigate in a microscopical transport model
the evolution of conical structures originating from the supersonic
projectile moving through the matter of ultrarelativistic particles.
Using different scenarios for the interaction between projectile and matter,
and different transport properties of the matter, we study the formation 
and structure of Mach cones. Furthermore, the two-particle correlations for
different viscosities are extracted from the numerical calculations and
we compare them to an analytical approximation. In addition, by adjusting
he cross section we investigate the influence of the viscosity to the
structure of Mach cones.
\end{abstract}
\PACS{PACS numbers come here}
  
\section{Introduction}

Highly energetic partons propagating through
the hot and dense QGP rapidly lose their energy and momentum as the energy
is deposited in the medium. Measurements of two- and three-particle
correlations in heavy-ion collisions show a complete
suppression of the away-side jet, whereas for lower $p_T$ a double
peak structure is observed  in the two-particle correlation
function \cite{Wang:2004kfa}. One possible and promising origin of these structures
is assumed to be the interaction of fast partons with the soft matter
which generates collective motion of the medium in form of Mach cones.
\cite{Stoecker:2004qu,Bouras:2010nt}.

For this purpose we investigate the propagation and formation of
Mach cones in the microscopic transport model BAMPS
(Boltzmann Approach of MultiParton Scatterings) \cite{Xu:2004mz}
in the limit of vanishing mass and very small shear viscosity over
entropy density ratio $\eta/s$ of the matter. Two different scenarios
for the jet are used. In addition, by adjusting $\eta/s$, the
influence of the viscosity on the profile of the Mach cone and
the corresponding two-particle correlation is explored for the
first time. The results presented are based on a recent publication
\cite{Bouras:2012mh}.

\section{Shock Waves and Mach cones}

\begin{figure*}[ht]
\centering 
\includegraphics[width=\textwidth]{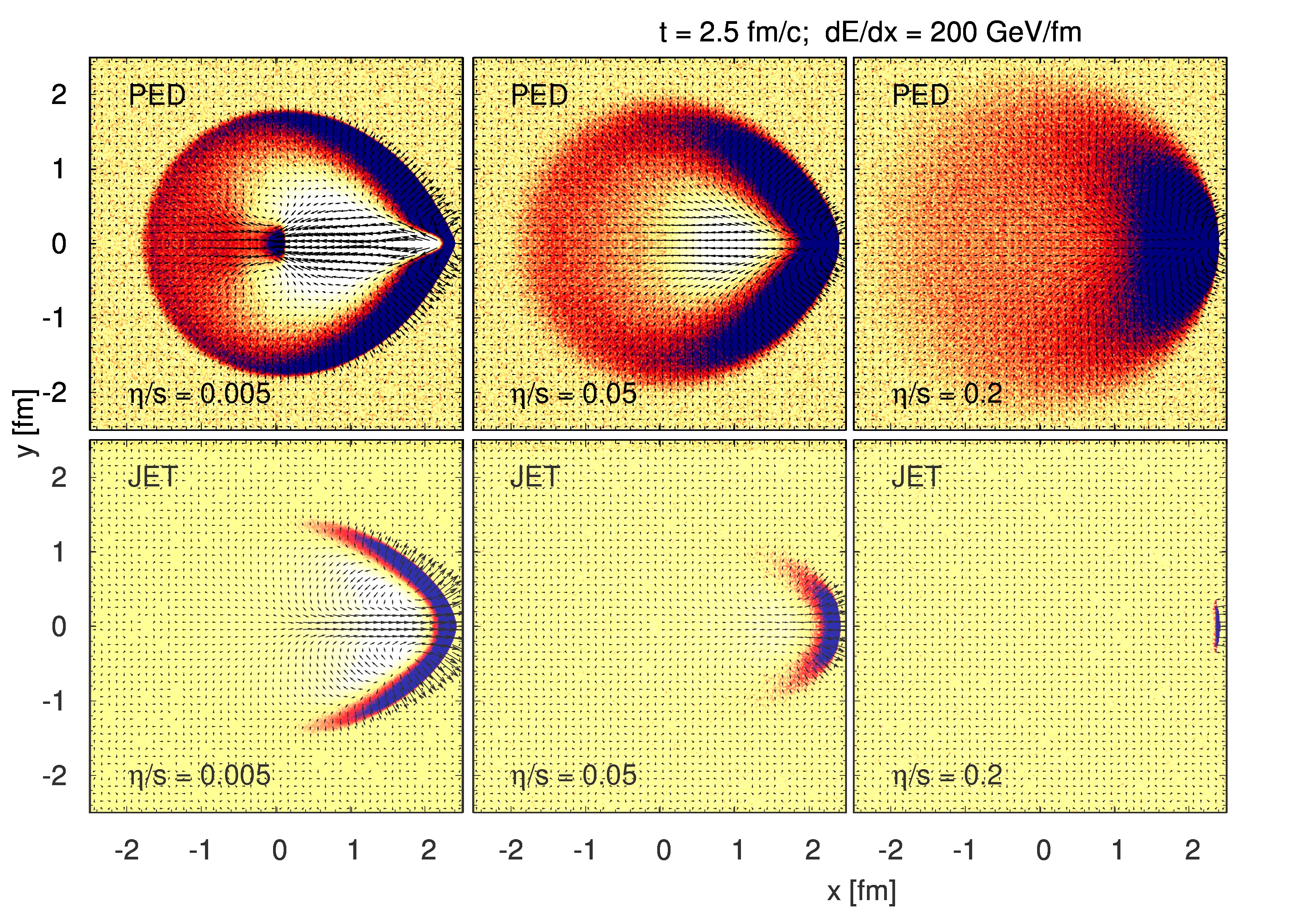}
\caption{(Color online) Transition from ideal to viscous Mach cones.
Shape of a Mach cone shown for different jet scenarios and different
viscosity over entropy density ratios, $\eta/s = 0.005$, $0.05$ and $0.5$.
The energy deposition is $dE/dx = 200$ GeV/fm. The upper
panel shows the pure energy deposition scenario (PED); the lower
panel shows the propagation of a highly energetic jet (JET) depositing
energy and momentum in $x$-direction. Depicted are the LRF energy density
within a specific range; as an overlay we show the velocity profile with
a scaled arrow length. The results are a snapshot of the
evolution at $t = 2.5$ fm/c.}
\label{fig:eDensity_viscDifSourceTerms}
\end{figure*}

Shock waves are phenomena which have their origin in
the collective motion of matter. In a simplified 
one-dimensional setup shock waves have already been studied 
within the framework of BAMPS for the perfect fluid limit
\cite{Bouras:2009nn,Bouras:2010hm}. Furthermore BAMPS
calculations have demonstrated  that the shock profile
is smeared out when viscosity is large.
It was also found that a clear observation of the shock within
the short time available in HIC requires a small viscosity.

In the following we study the evolution of "Mach cone"-like
structures with different scenarios of the jet-medium interaction
by using the parton cascade BAMPS. We focus on investigation
of Mach cone evolution in absence of any other effects
- i.e.\ we neglect such effects as initial fluctuations or
expansion, which are however relevant in HIC.
We use a static box with $T_{\rm med} = 400$ MeV and
binary collisions with an isotropic cross section.
Furthermore, we keep the mean free path $\lambda_{\rm mfp}$ of the medium
particles constant in all spatial cells by adjusting the
cross section according to $\sigma = 1 / (n\lambda_{\rm mfp})$,
where $n$ is the particle density. The related shear viscosity
for isotropic binary collisions is given by
$\eta = 0.4\,e \, \lambda_{\rm mfp}$ \cite{deGroot}.

\begin{figure}[h]
\centering 
\includegraphics[width=\columnwidth]{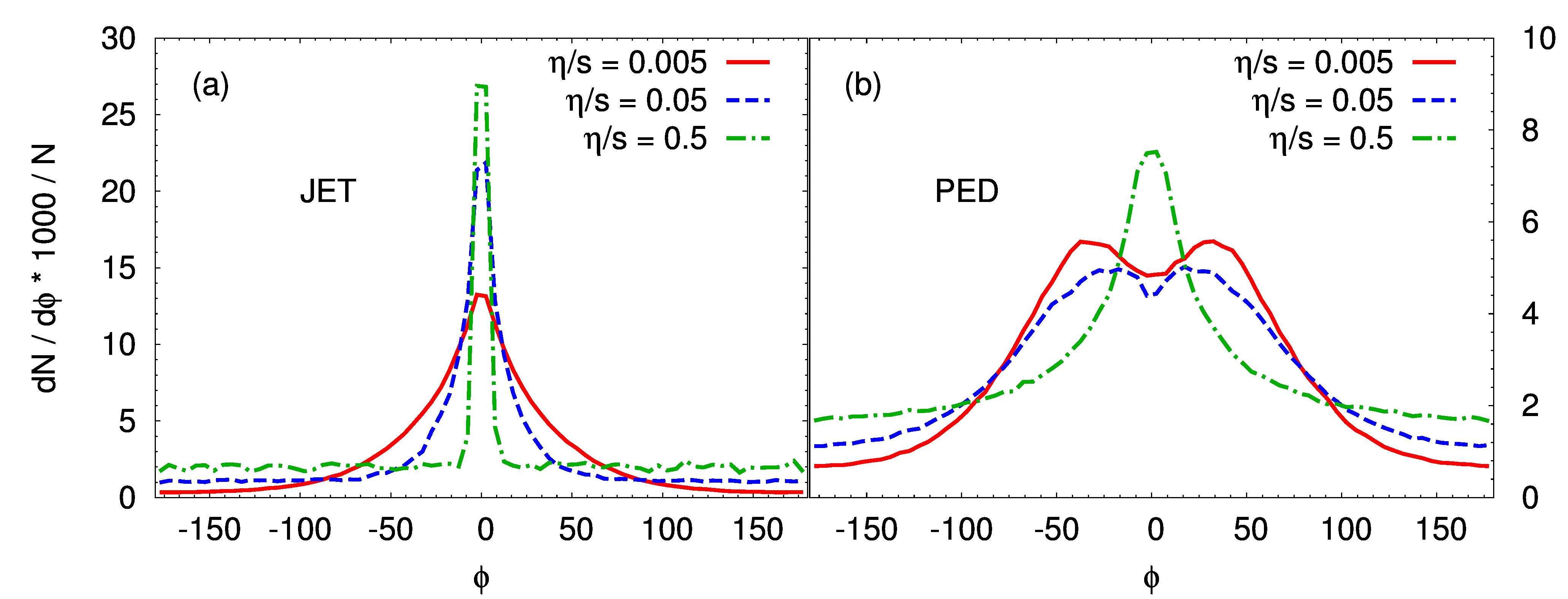}
\caption{(Color online) Two-particle correlations
$dN/(N d\phi)$ for different viscosities extracted from
calculations shown in Fig.~\ref{fig:eDensity_viscDifSourceTerms}.
The results are shown in the for the JET (a) and PED (b)
scenario for $dE/dx = 200$ GeV/fm.
}
\label{fig:num_visc_TPC}
\end{figure}

The Mach Cones studied here are induced by two different sources.
The first of them we refer to as the pure energy
deposition scenario (PED) \cite{Betz:2008ka}. This is simulated
by a moving source depositing momentum end energy isotropically
according to the thermal distribution $f(x,p) = exp(-E/T)$.
The second source we refer to as JET. This is simulated by a
highly massless particle (jet) which has only
momentum in $x$-direction, i.e. $p_{\rm x} = E_{ \rm jet}$.
After each timestep the energy of the jet particle is reset
to its initial value. For both scenarios the
sources are initialized at $t = 0$ fm/c at the position
$x = - 0.1$ fm and propagate in $x$-direction with
$v_{\rm source} = 1$, i.e. with the speed of light.

In Fig.~\ref{fig:eDensity_viscDifSourceTerms} we show the
Mach Cone structure for both PED scenario (upper panel) and
JET scenario (lower panel) with $\eta/s = 0.005$, $0.05$ and $0.5$
from left to right, respectively. We show a snapshot at $t = 2.5 \rm fm/c$.
The energy deposition rate is
fixed to $dE/dx = 200$ GeV/fm. In both scenarios, PED and JET,
for $\eta/s = 0.005$ (left panel), we observe a conical structure,
but with obvious differences. The PED case with the isotropic energy deposition
induces a spherical shock into back region; this structure is missing
in the JET scenario because of the high forward peaked momentum deposition.
Another difference is that in the JET scenario a clearly visible
head shock appears. This in turn is missing in the PED scenario.
Furthermore a (anti)-diffusion wake is induced by the JET (PED)
scenario. 

Adjusting the shear viscosity over entropy density ratio
$\eta/s = 0.05 - 0.5$ we observe a smearing out of the Mach
cone structure. For a sufficient high $\eta/s = 0.5$
the conical structure in both scenarios disappears.
This is true for shock fronts as well as for the
(anti-) diffusion wake. The difference between the PED and
the JET case is that as $\eta/s$ increases, in the PED
scenario the resulting "Mach cone" solution covers
approximately the same spatial region regardless of a value
of $\eta/s$, while in the JET case the structure is
concentrated more and more near the projectile as the
viscosity increases.

In Fig.~\ref{fig:num_visc_TPC} we show the two-particle correlations
extracted from BAMPS calculations of the Mach Cones shown in
Fig.~\ref{fig:eDensity_viscDifSourceTerms}. For the JET scenario (a)
and sufficiently small $\eta/s = 0.005$ we observe only a peak
in direction of the jet. The typical double peak structure, which
has been proposed as a possible signature of the Mach cone in HIC,
can only be observed for the PED scenario (b) and small $\eta/s$.
However, the PED scenario has no correspondence in heavy-ion physics.
For the JET scenario, which is a simplified model of jets depositing
energy and momentum, a double peak structure never appears. This is
due to the strong formation of a head shock and diffusion wake.

\section{Conclusions}

In summary, the evolution of Mach cones induced by two different
source terms, PED and JET, were investigated using a microscopic transport model.
The development of Mach cones is observed in case the viscosity of matter
is small enough. In addition, the effects of viscosity of the matter were
shown by adjusting the shear viscosity over entropy density ratio
$\eta/s$ from $0.005$ to $0.5$. A clear and unavoidable smearing of
the profile depending on a finite ratio of shear viscosity to entropy
density is clearly visible. Investigating the corresponding two-particle
correlations we see that Mach cones can not be connected
to double peak structures by any realistic picture of jets in HIC.

\section*{Acknowledgements}

The authors are grateful to the Center for the Scientific 
Computing (CSC) at Frankfurt for the computing resources.
This work was supported by the Helmholtz International Center
for FAIR within the framework of the LOEWE program 
launched by the State of Hesse.

\end{document}